\documentclass[conference]{IEEEtran}
\IEEEoverridecommandlockouts

\usepackage{cite}
\usepackage{amsmath,amssymb,amsfonts}
\usepackage{algorithmic}
\usepackage{graphicx}
\usepackage{textcomp}
\usepackage{xcolor}
\usepackage{multirow, threeparttable, booktabs, makecell}
\def\BibTeX{{\rm B\kern-.05em{\sc i\kern-.025em b}\kern-.08em
    T\kern-.1667em\lower.7ex\hbox{E}\kern-.125emX}}
\begin{document}
\title{Adaptive Data Augmentation with NaturalSpeech3 for Far-field Speaker Verification}

\author{\IEEEauthorblockN{1\textsuperscript{st} Li Zhang, 2\textsuperscript{nd} Jiyao Liu, 3\textsuperscript{nd} Lei Xie}
 
\IEEEauthorblockA{\textit{Audio, Speech and Language Processing Group (ASLP@NPU), School of Computer Science} \\
\textit{Northwestern Polytechnical University (NPU)}\\
Xi’an, China \\
850143245@mail.nwpu.edu.cn}
}

\maketitle

\begin{abstract}
The scarcity of speaker-annotated far-field speech presents a significant challenge in developing high-performance far-field speaker verification (SV) systems. While data augmentation using large-scale near-field speech has been a common strategy to address this limitation, the mismatch in acoustic environments between near-field and far-field speech significantly hinders the improvement of far-field SV effectiveness. In this paper, we propose an adaptive speech augmentation approach leveraging NaturalSpeech3, a pre-trained foundation text-to-speech (TTS) model, to convert near-field speech into far-field speech by incorporating far-field acoustic ambient noise for data augmentation. Specifically, we utilize FACodec from NaturalSpeech3 to decompose the speech waveform into distinct embedding subspaces —content, prosody, speaker, and residual (acoustic details) embeddings—and reconstruct the speech waveform from these disentangled representations. In our method, the prosody, content, and residual embeddings of far-field speech are combined with speaker embeddings from near-field speech to generate augmented pseudo far-field speech that maintains the speaker identity from the out-domain near-field speech while preserving the acoustic environment of the in-domain far-field speech. This approach not only serves as an effective strategy for augmenting training data for far-field speaker verification but also extends to cross-data augmentation for enrollment and test speech in evaluation trials. In augmentation of enrollment and test utterances, the method mitigates performance degradation caused by discrepancies in text content or environmental noise between enrollment and test data. This data augmentation method, which preserves the acoustic environment of the in-domain far-field data, qualifies as an adaptive augmentation method. Experimental results on FFSVC demonstrate that the adaptive data augmentation method significantly outperforms traditional approaches, such as random noise addition and reverberation, as well as other competitive data augmentation strategies.
\end{abstract}


\section{Introduction}
Speaker verification (SV) refers to the process of determining the acceptance or rejection of test utterances based on enrollment utterances \cite{naika2018overview}. In recent years, numerous speaker verification approaches leveraging deep learning techniques have exhibited remarkable recognition performance, particularly in near-field scenarios with less interference \cite{nagrani2020voxceleb,bu2017aishell,du2018aishell}.  
Nevertheless, their effects experience a substantial decline when confronted with speech collected from far-field environments \cite{huang2019intel, qin2022far}, predominantly attributable to the degradation of speech quality \cite{nandwana2019voices,qin2020hi} and paucity of speech in far-field scenarios \cite{movsner2022multisv,zhang2020npu}.

The degradation in speech quality primarily stems from three underlying factors. First, the speech signal is attenuated when propagating from the speaker to the microphones, resulting in low signal power and often also low signal-to-noise ratio (SNR) \cite{qin2020interspeech}. Secondly, in enclosed spaces like living or meeting rooms, the source signal reflects multiple times off walls and objects, causing multipath propagation, leading to temporal smearing of the source signal, known as reverberation \cite{qin2020hi}. Thirdly, there is a high probability of the microphone capturing various interfering sounds alongside the desired speech signal, such as the television or HVAC equipment \cite{qin2020hi}. These acoustic interferences are diverse and complex to simulate, difficult to simulate, and can significantly affect the performance of far-field loudspeaker verification. In addition, the scarcity of far-field speech in training is another hindrance to training superior far-field speaker verification systems.  Compared with the easy acquisition of near-field speech, the collection of far-field speech is more expensive and time-consuming compared with near-field speech collection \cite{qin2022far, du2018aishell}.

Many approaches have been proposed to solve the above problems and improve the performance of far-field speaker verification. Given these signal degradations, there are many front-end speech approaches introduced to far-field speaker verification systems \cite{movsner2022multi,liang2021multi,gao2022unet,movsner2018dereverberation}. Most front-end processing methods are borrowed from the speech enhancement field. The datasets used in experiments are released far-field speaker verification corpora \cite{qin2020interspeech,nandwana2019voices} or generated from manually simulated \cite{movsner2022multisv}. 
Data augmentation is the prevailing approach in response to the limited availability of data in far-field speaker verification. Conventional data augmentation methods typically involve performing transformations on the speech waveform, encompassing random additive noise \cite{qin2019far}, reverberation \cite{gusev2020stc}, variable speed \cite{zhang2022npu}, specAug \cite{park2019specaugment}. Nonetheless, these solutions need to consider the specific recording conditions of far-field speech, rendering them incapable of augmenting speech that conforms to the acoustic environment noise present in the application scenario. Typically, the augmentation parameters, such as the signal-to-noise ratio (SNR) and room size, are manually set, which requires a lot of trial-and-error and intuition \cite{lin2022robust}. However, a domain mismatch in the acoustic environment between near-field and far-field speech poses a significant limitation to enhancing the far-field speaker verification system \cite{zhang2023distance}.

To fill this research gap, this paper proposes an adaptive data augmentation approach to transform near-field speech into acoustic-matching far-field speech. We propose an innovative adaptive speech augmentation technique that utilizes NaturalSpeech3 \cite{ju2024naturalspeech}, a pre-trained universal text-to-speech (TTS) model, to transform near-field speech into far-field speech by integrating far-field acoustic ambient noise. This method leverages NaturalSpeech3 \cite{ju2024naturalspeech}'s FACodec, which disentangles the speech waveform into separate components—content, prosody, speaker, and residual embeddings (acoustic details)—and then reconstructs the waveform from these isolated elements. Our approach combines the prosody, content, and residual features of far-field speech with speaker embeddings from out-domain data. This combination generates augmented speech that retains the speaker's identity from the out-domain data while faithfully reproducing the acoustic characteristics of the in-domain far-field environment. This technique is highly effective for augmenting training datasets in far-field speaker verification tasks and can also be applied to cross-domain augmentation for enrollment and testing phases. This method reduces performance degradation by addressing inconsistencies in text content and environmental noise between enrollment and test data. Our approach, which preserves the acoustic characteristics of the in-domain far-field environment, stands out as an adaptive augmentation strategy.  

The contributions of this paper are as follows:
\begin{itemize}

  \item We propose an adaptive data augmentation method to simulate far-field speech with acoustic environment noise estimated from real recorded far-field speech named adaptive data augmentation.
 
  \item We leverage NaturalSpeech3's FACodec to disentangle and reconstruct speech components, ensuring the preservation of the acoustic environment. We group the non-speaker embedding from far-field speech with speaker embeddings from near-field speech to create an augmented speech that retains the speaker's identity from out-domain data while maintaining the acoustic characteristics of the in-domain far-field environment.
  \item The adaptive data augmentation effectively augments training data for far-field speaker verification and mitigates performance degradation by addressing text content and environmental noise discrepancies between enrollment and test data, making it a robust adaptive augmentation strategy.
  \item We assess the effectiveness of the adaptive data augmentation on far-field speaker verification. The results obtained from the evaluation of FFSVC2020 demonstrate the superiority of this method.
  
\end{itemize}
The remainder of the paper is structured as follows: Section II reviews related works, while Section III provides an overview of the adaptive data augmentation approach. Section IV details the experimental setup, followed by results and analysis in Section V. Finally, Section VI concludes the paper. 
\section{Related Work}


\subsection{Augmentation with Speech Transformations and Modifications}
Speech transformations and modifications are the easiest and most efficient ways to augment speech. The seminal work X-Vector \cite{snyder2018x} introduces adding noise and reverberation to the original speech as an inexpensive method to multiply the amount of training data and improve robustness. Recently, ECAPA-DTNN augmented with SpecAugment \cite{park2019specaugment,faisal2019specaugment} on frequency features, speed perturber \cite{geng2022investigation} on raw waveforms achieves state-of-art results on VoxCeleb datasets. The above transformations and modifications of speech are applied to far-field speaker verification. Almost all the winners of recent far-field speaker verification competitions \cite{qin2022far,qin2020interspeech,nandwana2019voices} have used the above-mentioned speech augmentation schemes. In addition, time-domain random masking is demonstrated effectiveness in robust speaker verification \cite{lian2021improved,zhang2022npu}.  In the realm of speaker verification, a wide array of techniques, such as introducing random noise and reverberation, applying SpecAugmentation, employing speed perturbation, and implementing time-domain speech random masking, are commonly integrated into the training processes of most speaker verification models.
\subsection{Augmentation with Near-field Speech}
Acquiring labeled far-field speech data for far-field speaker verification necessitates a more significant investment of time and financial resources than obtaining near-field speech \cite{qin2020interspeech, du2018aishell}. As a result, it is customary to exploit extensive open-source repositories of near-field speech datasets to improve the performance of far-field speaker verification systems \cite{zhang2020npu,qin2019far,ko2017study}.
Nonetheless, owing to the inherent divergence between the acoustical domains of far-field and near-field speech, the direct amalgamation of these two types of speech data yields generally constrained outcomes. When precise information regarding the far-field acoustic capture environment and microphone array configuration is available, it becomes feasible to artificially transform near-field speech into synthetic far-field speech that emulates the equivalent acoustic environmental noise profile of genuine far-field speech. However, obtaining detailed recording scene configurations of far-field training speech in real life is complicated. The majority of speech augmentation methodologies entail applying randomized transformations and alterations to both near-field and far-field speech, thereby generating a varied and extensive set of training samples for far-field speaker verification. The blind random augmentation, lacking consideration for specific far-field speech recording conditions, results in minimal enhancement of speaker verification performance within particular far-field scenarios.

\subsection{Augmentation from Generative Models}
Several works use generative models for speaker verification data augmentation. Du et al. \cite{du2021synaug} propose a synthesis-based data augmentation method (SynAug) to expand the training set with more speakers and text-controlled synthesized speech. Bird et al. \cite{bird2020overcoming} are tackling the challenge of data scarcity in speaker identification by enhancing datasets by incorporating synthetic MFCCs via a character-level RNN methodology. In addition, Wang et al. \cite{9167416,wu19e_interspeech} propose variational autoencoder (VAE) and generative adversarial network (GAN) to generate speaker embeddings for back-end PLDA training. Recently, Hu et al. \cite{10094698} employed a generative model based on StarGAN to learn cross-domain mappings from single-speaker multi-condition speech, enabling the generation of absent domain data for all speakers, thereby augmenting the training set's intra-class diversity. While this voice conversion technique mitigates the domain discrepancy among various data domains, it does not augment the speaker count within the dataset or generate additional transformed speech corresponding to the ambient noise of the present environment.

While the data above augmentation strategies have demonstrated significant effectiveness, there still needs to be a gap in preserving the acoustic environment and exploring adaptive data augmentation within test trials. We propose an adaptive data augmentation approach for far-field speaker verification to address this shortfall, utilizing the NaturalSpeech3 \cite{ju2024naturalspeech} TTS model. This method aims to mitigate mismatch issues within the training or testing sets, thereby enhancing the robustness of the verification process.
\section{Methods}
The adaptive data augmentation framework using NaturalSpeech3 \cite{ju2024naturalspeech} is illustrated in Fig.\ ref{fig:overview}. The framework comprises three core modules: a disentanglement module, a voice conversion module, and a speaker encoder module. The disentanglement module, FACodec from NaturalSpeech3 \cite{ju2024naturalspeech}, decomposes the speech waveform into distinct subspaces, including content, prosody, speaker, and residual (acoustic details) embeddings. The voice conversion module facilitates the transfer of speaker identity from far-field speech to the timbre of near-field speech. The speaker encoder, a conventional speaker recognition model, is employed to classify speaker identities using both the augmented pseudo far-field speech and real far-field speech. This adaptive data augmentation method preserves the acoustic characteristics and textual content of the far-field data while integrating the speaker identity from the near-field data. The resulting augmented data effectively enhances the diversity of speaker identities in the training set by incorporating all speaker identities from the near-field data. This increases the model's generalization capability across various speaker identities, thereby improving its robustness. Moreover, the joint training of the speaker encoder module using both voice-converted data and original far-field data further enhances the performance of the SV system.

Suppose the far-field speech dataset is denoted as $X^f = \{x^f_1, x^f_2, x^f_3, \dots, x^f_p\}$ with corresponding speaker labels $Y^f = \{y^f_1, y^f_2, y^f_3, \dots, y^f_q\}$, where $q$ represents the number of unique speakers in the far-field dataset. Similarly, the near-field speech dataset is denoted as $X^n = \{x^n_1, x^n_2, x^n_3, \dots, x^n_s\}$ with speaker labels $Y^n = \{y^n_1, y^n_2, y^n_3, \dots, y^n_t\}$, where $t$ represents the number of unique speakers in the near-field dataset. By converting the speaker identities from $X^f$ into the speaker identities of $Y^n$, we can potentially generate up to $q \times t$ pseudo far-field speaker identities. We shuffle and select data corresponding to $t$ speakers from this pool.  Consequently, the combined training set, which includes both the pseudo far-field data and the real far-field data, will encompass speaker labels from $q + t$ unique identities. This expanded training set increases the diversity of speaker identities, thereby enhancing the model's ability to generalize across different speaker identities and improving its robustness in speaker verification tasks.
 
\begin{figure}[th]
\centering
\centerline{\includegraphics[width=0.5\textwidth]{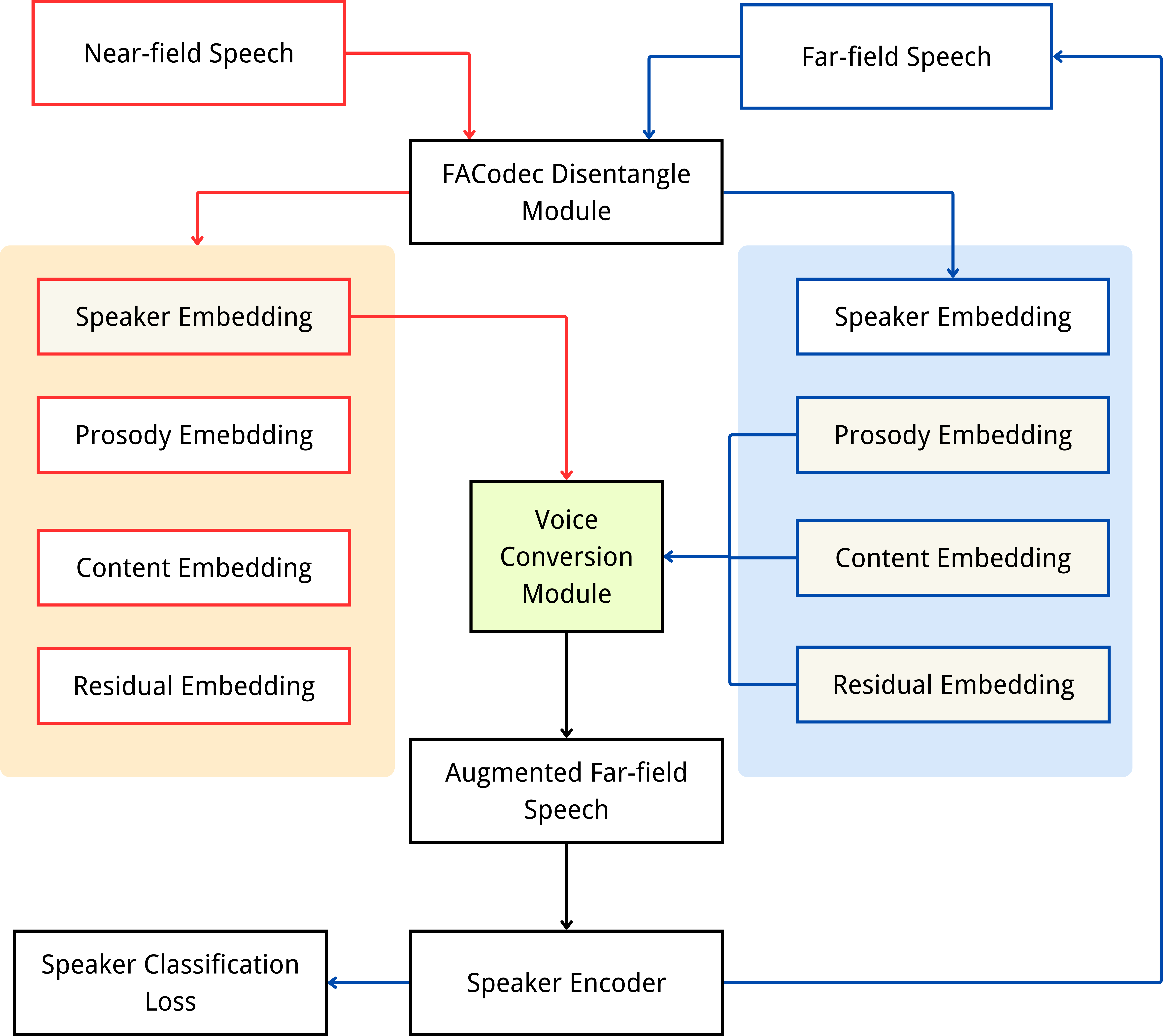}}
\caption{ The Overview of the Adaptive Data Augmentation.  }
\label{fig:overview}
\end{figure}

\subsection{Revisiting NaturalSpeech3 }\label{AA}
NaturalSpeech3 \cite{ju2024naturalspeech} is a cutting-edge large-scale TTS speech foundation model that addresses the challenges of generating high-quality speech with precise similarity and prosody by employing a novel factorization approach. The core idea of NaturalSpeech3 \cite{ju2024naturalspeech} involves decomposing speech into distinct subspaces—content, prosody, timbre, and acoustic details—and generating each attribute separately. This factorized strategy simplifies the complex task of speech modeling, leading to improved speech generation. The system's key components include the Factorized Vector Quantization (FVQ) Codec, known as FACodec, which effectively disentangles these subspaces using advanced techniques such as information bottlenecks, supervised losses, and adversarial training. Additionally, the Factorized Diffusion Model generates speech representations for each attribute based on specific prompts, allowing for independent control of different speech characteristics. This design results in several benefits: efficient learning and high-quality speech generation, simplified modeling of timbre, and enhanced controllability through prompt-based adjustments. Experiments have shown that NaturalSpeech3 \cite{ju2024naturalspeech} outperforms state-of-the-art TTS systems in terms of quality, similarity, prosody, and intelligibility, achieving a level of quality comparable to human recordings. Moreover, the system's scalability, with 1 billion parameters and training on 200,000 hours of data, further highlights its robustness and effectiveness.

In this paper, we employ the FACodec $\mathcal{D}_{facodec}$ component of NaturalSpeech3 \cite{ju2024naturalspeech} to effectively disentangle raw audio into distinct embedding subspaces, which disentangle the speech into prosody, content, speaker identity, and residual embeddings. This process is mathematically formulated as follows:

\begin{equation}
  F^f_{p}, F^f_{c}, F^f_{s}, F^f_{r} =  \mathcal{D}_{facodec}(X^f) 
\end{equation}

Here, $F_f^{{p}}$, $F_f^{{c}}$, $F_f^{{s}}$, and $F_f^{{r}}$ represent the prosody, content, speaker identity, and residual embeddings derived from the far-field speech $X^f$, respectively. 

Similarly, for near-field speech, the disentanglement process can be expressed as:

\begin{equation}
  F^n_{{p}}, F^n_{{c}}, F^n_{{s}}, F^n_{{r}} =  \mathcal{D}_{facodec}(X^n) 
\end{equation}

In the above formulas, $ F^n_{{p}}, F^n_{{c}}, F^n_{{s}}$ correspond to the prosody, content, speaker identity, and residual embeddings of the near-field speech $X^n$, respectively. By disentangling these embedding subspaces, FACodec allows for more granular manipulation and generation of speech attributes, thereby enabling a more flexible and precise text-to-speech (TTS) process. 

In our adaptive data augmentation method, we effectively utilize the disentanglement of speech embeddings to separate and analyze the distinct attributes of both far-field and near-field speech. This process sets the foundation for combining the non-speaker embeddings from the far-field speech with the speaker embeddings from the near-field speech, facilitating the synthesis of new pseudo far-field speech data. This synthesized data is then employed to enhance the robustness and generalization of the speaker verification system.

\subsection{Voice Conversion}

With the disentangled embeddings of the prosody, content, speaker identity, and residual embeddings, we use the voice conversion module in NaturalSpeech3 \cite{ju2024naturalspeech} to converse the speaker identities of far-field speech into the speaker identities of near-field speech. The voice conversion module in NaturalSpeech3 \cite{ju2024naturalspeech} builds on the disentangled embeddings obtained from FACodec, which separates prosody, content, speaker identity, and residual features for both far-field and near-field speech. This module is designed to modify the speaker identity of far-field speech by replacing it with near-field speech while preserving the other attributes, such as prosody, content, and residual features. The process can be mathematically expressed as follows:

\[
S_{pseudo-far} = \mathcal{C}(F^f_{p}, F^f_{c}, F^n_{s}, F^f_{r})
\]

Here, $S_{pesudo-far}$ represents the converted pseudo far-field speech that combines the prosody, content, and residual embeddings from the far-field speech with the speaker identity embedding from the near-field speech. The module first selects the relevant embeddings—prosody, content, and residuals—from the far-field speech and the speaker identity from the near-field speech. It then substitutes the far-field speaker identity \( F^f_{\text{s}} \) with the near-field speaker identity \( F^n_{\text{s}} \), ensuring that the new speech reflects the vocal characteristics of the near-field speaker while retaining the prosody and content of the far-field speech. 
This voice conversion mechanism is particularly valuable in applications such as SV tasks, speech synthesis, and other TTS tasks, where it is crucial to adapt the speaker identity while maintaining key acoustic and prosodic elements.
\subsection{Speaker Encoder}
We utilize both the augmented pseudo far-field speech data and the real far-field speech to train the speaker encoder to improve the generalization and robustness of the speaker model in far-field SV tasks. The speaker encoder model employed in this paper is based on emphasized channel attention, propagation and aggregation time delay neural network (ECAPA-TDNN) \cite{desplanques2020ecapa} architecture with 1024 channels, referred to as ECAPA-TDNN (1024). 
ECAPA-TDNN is a deep neural network model for speaker verification that enhances traditional TDNN by incorporating channel attention and multi-scale feature aggregation to improve speaker discrimination in complex acoustic environments.The framework of the speaker encoder comprises two primary components: the speaker embedding extractor, denoted as \( F_E \), and the classifier, denoted as \( F_C \). The training process utilizes the additive angular margin softmax \cite{deng2019arcface} (AAM-Softmax) loss function, which is particularly effective for speaker verification tasks by enhancing the separation between different speaker classes.

The speaker labels for the training data are divided into two sets:  $Y^f = \{y^f_1, y^f_2, y^f_3, \dots, y^f_q\}$ representing the far-field speakers, and  $Y^n = \{y^n_1, y^n_2, y^s_3, \dots, y^n_q\}$ representing the pseudo far-field speaker labels ( original near-field speakers). The total number of speakers is \( q + t \), which determines the number of neurons in the classification layer. The speaker prediction error during training, denoted as \( \mathcal{L}(f_C, f_E) \), is evaluated using the speaker classification loss, which is based on the AAM-Softmax function \cite{deng2019arcface}. 
\section{Experimental Setup}
\subsection{Dataset}
In this study, we utilize the FFSVC 2020 \cite{qin2020interspeech} dataset as the far-field data and the AISHELL-2 dataset \cite{du2018aishell} as the near-field data. To expand the training set of the far-field data, we transfer the speaker identities from the near-field data (AISHELL-2) to the speaker labels of the far-field data (FFSVC2020).

\textbf{FFSVC2020:} The FFSVC2020 dataset includes a training partition with 120 speakers and a development partition with 35 speakers. The evaluation phase consists of test trials across three distinct tasks: FFSVC-Task1, which focuses on far-field text-dependent speaker verification using a single microphone array; FFSVC2020-Task2, which addresses far-field text-independent speaker verification using a single microphone array; and FFSVC2020-Task3, which involves far-field text-dependent speaker verification using distributed microphone arrays. This setup allows for comprehensive evaluation across different verification scenarios, contributing to the robustness of the speaker verification system.

\textbf{AISHELL-2:} The AISHELL-2 corpus, a 1,000-hour clean read speech dataset recorded with a high-fidelity microphone. The dataset features 1,991 speakers (845 male, 1,146 female) aged 11 to 40+, with various accents—1,293 Northern, 678 Southern, and 20 other accents. Recordings occurred in two settings: 1,347 in a studio and the rest in a living room with natural reverberation. The content covers eight topics, including voice commands, places of interest, entertainment, finance, technology, sports, English spellings, and free speech. 

\subsection{Implement Details}

In this paper, the speaker encoder model employed is the ECAPA-TDANN (1024) \cite{desplanques2020ecapa}, utilizing the additive angular margin softmax (AAM-softmax) loss function \cite{deng2019arcface} with a margin of 0.2 and a scale factor of 30. The speaker embedding models are trained using 80-dimensional log Mel-filter bank features, with a window size of 25 ms and a window shift of 10 ms.
\begin{table*}[th]
\caption{  Comparison Results EER (\%)/minDCF (p=0.01) on FFSVC2020}
\label{tab:results1}
{\fontsize{4pt}{5.3pt}\selectfont 
\setlength\arrayrulewidth{0.1pt} 
\setlength{\heavyrulewidth}{0.1pt} 
\setlength{\aboverulesep}{0.0pt} 
\setlength{\belowrulesep}{0.0pt} 
\label{tab:ffsvc_results1}
\resizebox{\linewidth}{!}{\begin{tabular}{lcccccc}
\hline
\multicolumn{1}{c}{\multirow{2}{*}{Method}} & \multicolumn{2}{c}{FFSVC2020-Task1}                    & \multicolumn{2}{c}{FFSVC2020-Task2}                    & \multicolumn{2}{c}{FFSVC2020-Task3}                    \\ \cline{2-7} 
\multicolumn{1}{c}{}                        & EER                       & minDCF                     & EER                       & minDCF                     & EER                       & minDCF                     \\ \hline
Baseline\cite{desplanques2020ecapa}                                   & 10.00                     & 0.8399                     & 14.06                     & 0.9575                     & 8.711                     & 0.8203                     \\ \cline{2-7} 
In-domain Data Augmentation \cite{snyder2015musan,habets2006room} & 7.224                     & 0.7325                     & 12.81                     & 0.9228                     & 6.548                     & 0.7684                     \\ \cline{2-7} 
Out-domain Data Augmentation \cite{nagrani2020voxceleb}               & 7.429                     & 0.7376                      & 8.128                     & 0.7434                     & 6.610                     & 0.6970                     \\ \cline{2-7} 
SpecAug \cite{park2019specaugment}                                   & 8.082                     & 0.7443                     & 9.737                     & 0.8477                     & 7.120                     & 0.6824                     \\ \cline{2-7} 
Speech Perturbation \cite{yamamoto2019speaker}                              & 9.061                     & 0.7576                     & 11.21                     & 0.9012                     & 6.329                     & 0.7359                     \\ \cline{2-7} 
Shuffle Augmentation \cite{sato2023shuffleaugment}                         & 8.694                     & 0.8209                     & 10.04                     & 0.8386                     & 7.568                     & 0.7826                     \\ \cline{2-7} 
Filter Augmentation \cite{nam2022filteraugment}                         & \multicolumn{1}{l}{7.029} & \multicolumn{1}{l}{0.6656} & \multicolumn{1}{l}{8.898} & \multicolumn{1}{l}{0.8443} & \multicolumn{1}{l}{7.102} & \multicolumn{1}{l}{0.7325} \\ \cline{2-7} 
VC2Aug \cite{hu2023stargan}                                     & \multicolumn{1}{l}{6.990} & \multicolumn{1}{l}{0.6415} & \multicolumn{1}{l}{8.121} & \multicolumn{1}{l}{0.8192} & \multicolumn{1}{l}{6.638} & \multicolumn{1}{l}{0.6943} \\ \cline{2-7} 
\textbf{Adaptive Data Augmentation~(Train)}                 & \textbf{6.250 }                    & \textbf{0.5866}                     & \textbf{7.932}                     & \textbf{0.7928}                     & \textbf{5.836}                     & \textbf{0.6012}                     \\ \cline{2-7} 
\textbf{Adaptive Data Augmentation~(Test)}                 & \textbf{6.022 }                    & \textbf{0.5366}                     & \textbf{7.669}                     & \textbf{0.6892}                     & \textbf{5.707}                     & \textbf{0.5820}                     \\ \hline
 
\end{tabular}}}
\end{table*}
\subsection{Comparison Methods}

We conduct a comprehensive comparison of the adaptive data augmentation method with several alternative data augmentation techniques in the context of SV tasks. The methods compared are outlined as follows:

\begin{itemize}

\item In-domain Data Augmentation: Noise, music, and babble sounds from the MUSAN dataset \cite{snyder2015musan} are added to the original FFSVC2020 training set speech. Additionally, reverberation is applied using room impulse responses (RIRs) from the RIR noise dataset \cite{habets2006room}. 

\item Out-domain Data Augmentation: The VoxCeleb 1 \& 2 datasets \cite{nagrani2020voxceleb} are utilized as out-domain data to enhance performance on the FFSVC2020 far-field SV task.

\item SpecAugment: Time and frequency masking, along with time warping, are applied to the input spectrum (frequency-domain implementation) \cite{park2019specaugment}.

\item Speech Perturbation: Using the Speaker-Aug method \cite{yamamoto2019speaker} via the SOX toolkit, speech speed is altered to 0.9 or 1.1 times the original, generating additional utterances. This increases the speaker count to 360 by tripling the original 120 speakers.

\item Shuffle Augmentation: This method randomizes the time order of an input sequence at irrelevant time scales, generating multiple variants without compromising critical temporal information \cite{sato2023shuffleaugment}.

\item Filter Augmentation: This method simulates acoustic filters by applying varying weights across frequency bands, enhancing the model's ability to extract pertinent information from a broader frequency range \cite{nam2022filteraugment}.

\item VC2Aug: VC2Aug utilizes a StarGAN-based voice conversion model to create cross-domain mappings from single-speaker multi-condition data, generating missing domain data for all speakers, thereby increasing intra-class diversity within the training set \cite{hu2023stargan}.

\end{itemize}

\subsection{Score Metric}
In the test phase, we use cosine similarity as the scoring criterion. The performance metrics are equal error rate (EER) and minimum detection cost function (minDCF) [27], which is evaluated with $P_{\text {target }}=0.01, C_{\text {miss }}=C_{f a}=1$.
 
\section{Experimental Results}
\subsection{Comparison Results on FFSVC2020}\label{AAA}
The experimental results, as shown in TABLE \ref{tab:results1}, highlight significant performance improvements across all tasks on the FFSVC2020 dataset when various data augmentation techniques are applied, in comparison to the baseline model. The baseline model consistently displays the highest EER and minDCF values across all tasks, with EERs of 10.00\% for FFSVC2020-Task1, 14.06\% for FFSVC2020-Task2, and 8.711\% for FFSVC2020-Task3, and corresponding minDCF values of 0.8399, 0.9575, and 0.8203, respectively.

When different data augmentation methods are employed, substantial reductions in both EER and minDCF are observed, reflecting the effectiveness of these techniques in enhancing model performance. For FFSVC2020-Task1, the application of data augmentation results in a reduction in EER by 25.71\% to 37.5\% and a decrease in minDCF by 8.04\% to 30.17\%, indicating a significant improvement in the system's ability to accurately verify speakers.

Similarly, FFSVC2020-Task2 benefits greatly from data augmentation, with EER reductions ranging from 8.77\% to 43.6\% and minDCF reductions from 3.64\% to 17.21\%. These improvements suggest that augmented data assists the system in better managing the complexities of speaker verification in more challenging scenarios, leading to more accurate and reliable outcomes.

Task 3, which has the lowest baseline EER and minDCF values, also shows notable improvements with data augmentation. EER decreases by 3.57\% to 33.03\%, while minDCF is reduced by 2.76\% to 26.72\%. Although the baseline performance in this task is relatively strong, the results demonstrate that data augmentation can further enhance the system's robustness and accuracy, even when initial performance is already high.

The introduction of Adaptive Data Augmentation (Train) is particularly noteworthy, as it yields the lowest EER and minDCF values across all tasks. Specifically, this method reduces EER by 37.5\% for FFSVC2020-Task1, 43.6\% for FFSVC2020-Task2, and 33.03\% for FFSVC2020-Task3, with corresponding minDCF reductions of 30.17\%, 17.21\%, and 26.72\%. These significant improvements underscore the importance of well-designed data augmentation strategies in boosting speaker verification performance, especially under far-field conditions where the challenges of accurate verification are more pronounced.

The final row of the TABLE \ref{tab:results1} introduces Adaptive Data Augmentation (Test), a new approach that further improves system performance by addressing the mismatch between enrollment and test conditions. In this method, the speaker identity from the enrollment set is converted into the speaker identity of the test set (and vice versa), generating augmented samples for testing. This process mitigates the mismatch issues between enrollment and test data, leading to even further reductions in EER and minDCF. For FFSVC2020-Task1, the EER is reduced to 6.022\% with a minDCF of 0.5366, while FFSVC2020-Task2 sees an EER of 7.669\% and a minDCF of 0.6892. FFSVC2020-Task3 achieves the lowest EER of 5.707\% and a minDCF of 0.5820 with this approach.
\begin{figure}[th]
\centering
\centerline{\includegraphics[width=0.4\textwidth]{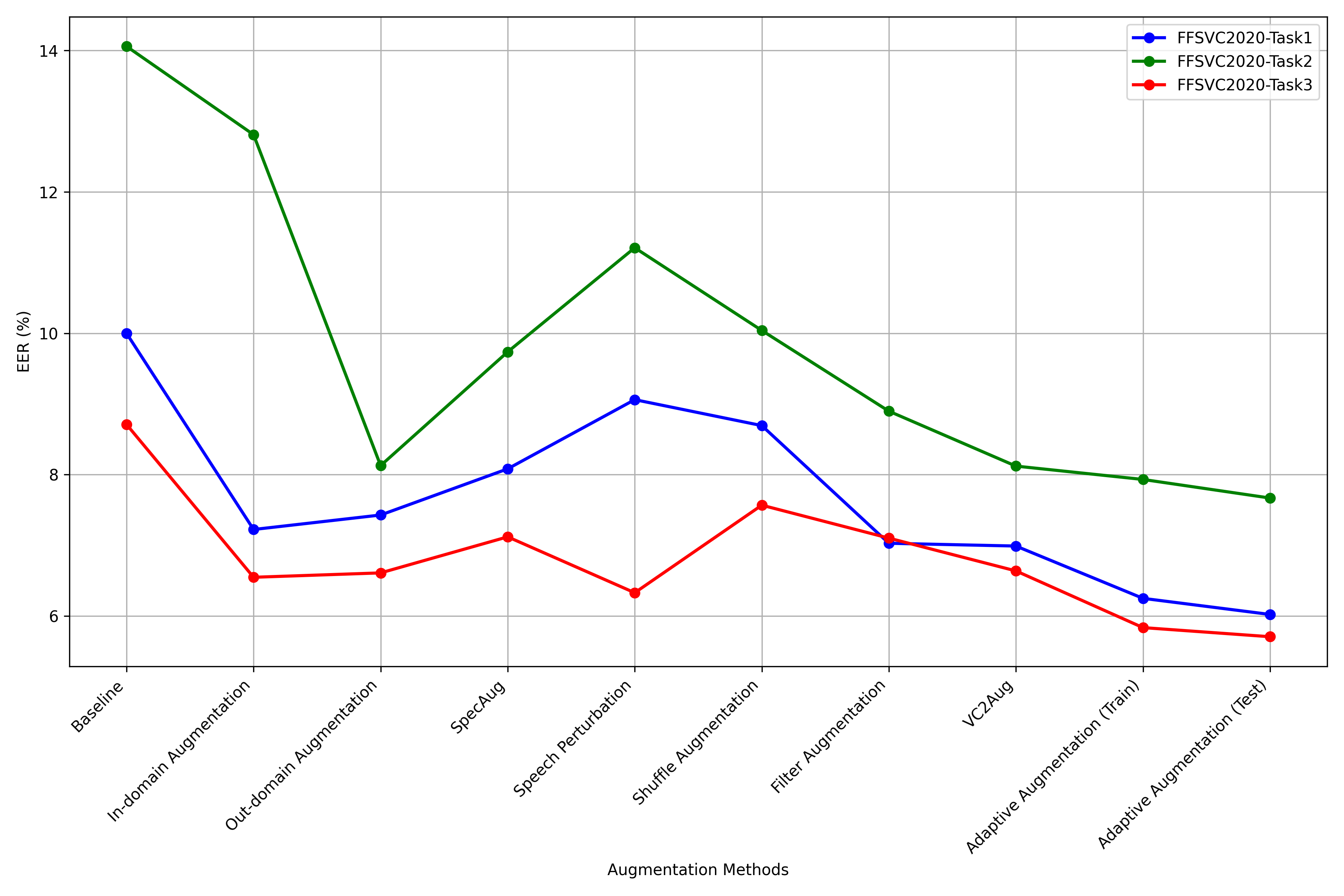}}
\caption{EER Comparison Across FFSVC2020 Tasks for Different Augmentation Methods}
\label{fig:EER_Comparison_FFSVC2020}
\end{figure}

To further visualize the trends in EER across different augmentation methods, we present a line graph of the EER values for each method, as shown in Fig. \ref{fig:EER_Comparison_FFSVC2020}. The visualization highlights the critical role of data augmentation in enhancing the effectiveness of speaker verification systems. By significantly reducing errors and improving detection performance, these techniques make the systems more robust and reliable, particularly in far-field scenarios where acoustic conditions pose additional challenges. The integration of adaptive data augmentation methods, as demonstrated by the improvements in both the training and testing phases, is essential for achieving superior performance in SV tasks.


\subsection{Augmented Sample Adaptive Analysis}\label{ITH}
\begin{figure*}[th]
\centering
\centerline{\includegraphics[width=0.74\textwidth]{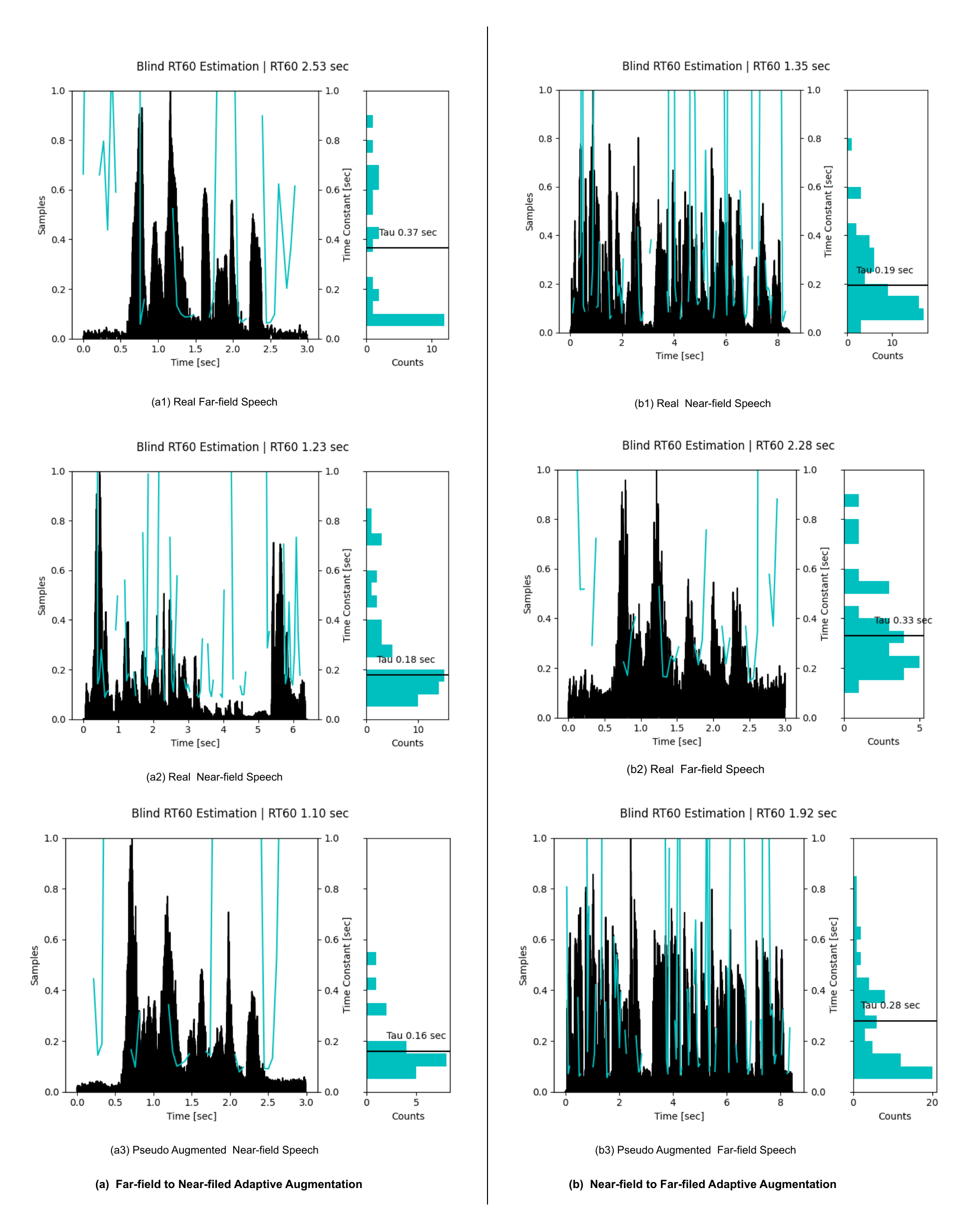}}
\caption{Visualization of Estimated RT60 for Real Speech and Adaptively Augmented Speech.}
\label{fig:visual}
\end{figure*}
To evaluate whether our augmented samples demonstrate adaptability—specifically, their ability to enable near-field data to simulate the acoustic characteristics of the current far-field environment—we employed a Blind RT60 Estimator \cite{ratnam2003blind}. This estimator was used to calculate the RT60 values of both the far-field and synthesized pseudo-far-field audio. Comparable RT60 values generally indicate that the two audio samples are recorded in similar acoustic environments, potentially in the same or comparable rooms. Longer RT60 values typically suggest that the audio is recorded in larger or more reflective spaces, such as an open hall or a room with limited sound absorption. In contrast, shorter RT60 values are indicative of smaller or acoustically treated spaces.

Our analysis confirms that the synthesized speech effectively adapts to both the training and testing environments, mitigating potential mismatch issues arising from data augmentation. The following visualization compares the estimated RT60 values for the far-field audio and the synthesized pseudo-far-field audio:

As shown in Fig. \ref{fig:visual}, the RT60 values of the augmented speech align more closely with those of the original far-field audio, even when the speech is converted from far-field to near-field. Conversely, when augmenting near-field speech using the speaker’s identity from the far-field, the RT60 values of the augmented data are closer to those of the near-field speech. This alignment suggests that the augmented speech captures the original environment's acoustic properties more accurately than those of the target speaker’s environment.
\section{Conclusion}

In this paper, we propose an adaptive data augmentation method with NaturalSpeech3 \cite{ju2024naturalspeech} for far-field speaker verification tasks. The proposed adaptive speech augmentation approach leveraging NaturalSpeech3 \cite{ju2024naturalspeech} effectively addresses the challenges posed by the scarcity of speaker-annotated far-field speech in developing high-performance far-field speaker verification systems. By converting near-field speech into far-field speech by incorporating far-field acoustic ambient noise and utilizing the FACodec to disentangle and reconstruct speech components, the method preserves the acoustic environment of the in-domain data while maintaining speaker identity. This method not only enhances training data augmentation but also improves cross-data augmentation for enrollment and test phases, mitigating performance degradation due to environmental discrepancies. Experimental results confirm that this adaptive augmentation method significantly outperforms traditional approaches and other competitive strategies, making it a valuable tool for advancing far-field speaker verification systems.

\bibliographystyle{unsrt}
\bibliography{refs}
\end{document}